\documentclass{jpsj-suppl}
\usepackage{txfonts} 
\usepackage{bm,color}

\renewcommand{\theenumi}{\roman{enumi}}

\title{Ingredients of nuclear matrix element \\ for two-neutrino double-beta decay of $^{48}$Ca}

\author{Y. \textsc{Iwata}$^{1}$, N. \textsc{Shimizu}$^{1}$, Y. \textsc{Utsuno}$^{2}$, M. \textsc{Honma}$^{3}$, T. \textsc{Abe}$^{4}$, and T. \textsc{Otsuka}$^{1,4,5}$}

\inst{$^{1}$Center for Nuclear Study, School of Science, The University of Tokyo, Japan \\
$^{2}$Japan Atomic Energy Agency, Japan \\
$^{3}$Center for Mathematical Sciences, University of Aizu, Japan \\
$^{4}$Department of Physics, School of Science, The University of Tokyo, Japan\\
$^{5}$National Superconducting Cyclotron Laboratory, Michigan State University, USA }

\email{iwata@cns.s.u-tokyo.ac.jp}

\recdate{July 31, 2014}

\abst{Large-scale shell model calculations including two major shells are carried out, and the ingredients of nuclear matrix element for two-neutrino double beta decay are investigated. 
Based on the comparison between the shell model calculations accounting only for one major shell ($pf$-shell) and those for two major shells ($sdpf$-shell),  the effect due to the excitation across the two major shells is quantitatively evaluated.}

\kword{Double beta decay, Nuclear matrix element, Shell model calculation}

\begin{document}
\maketitle

\section{Introduction}
There are two types of double-beta decay processes depending on whether neutrinos are emitted or not.
The former one is referred to two-neutrino double-beta decay, and the latter one to neutrino-less double beta decay.
Only the two-neutrino double-beta decay is admissible if the neutrino is the Dirac
particle, while both types can take place if the neutrino is
the Majorana particle.
In this sense experimental observation of neutrino-less double beta
process has an impact on determining one of the most fundamental properties of neutrino.

\begin{figure} [t]
 \includegraphics[width=15.0cm]{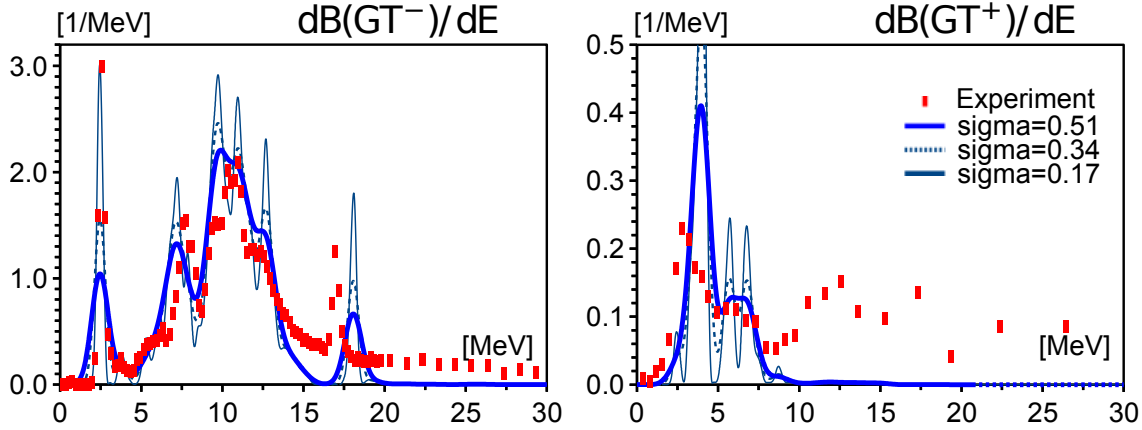}
\caption{(color online) GT$^{\pm}$ transition strength based on a shell
  model calculation (blue curves) accounting only for one major shell (GXPF1A) and experimental data (red bars)~\cite{09yako}.
Horizontal axis means the excitation energy of $1^+_c$ state of $^{48}$Sc measured from the ground state of $^{48}$Sc (see Eq. \eqref{mat1}).
The Gaussian with its square root of the variance: $\sigma$= 0.170, 0.340 and 0.510~MeV (corresponding to thin one, dotted one, and thick one, respectively) is applied to make comparable theoretical curves to the experiment. 
GT$^-$ transition strength from $^{48}$Ca to $^{48}$Sc ($\sigma=$0.085~MeV corresponds to the full width at half maximum of experimental energy resolution 0.2~MeV) is shown in the left panel, and GT$^+$ transition strength from $^{48}$Ti to $^{48}$Sc  ($\sigma=$0.170~MeV corresponds to the full width at half maximum of experimental energy resolution 0.4~MeV) is shown in the right panel.
The experimental data may or may not include isovector spin monopole transition strength~\cite{yakonotice} in addition to the Gamow-Teller transition strength.} 
\label{fig1}
\end{figure}

In this article ingredients of the nuclear matrix element for two-neutrino double-beta decay of $^{48}$Ca are calculated based on large-scale shell model calculations (KSHELL \cite{13shimizu}).
Two-neutrino double-beta decay processes consist mostly of double Gamow-Teller transition processes, so that two kinds of experiments are associated with the two-neutrino double-beta decay; one is the double-beta-decay half-life experiment, and the other is the Gamow-Teller transition experiments.
The half-life was measured to be $(4.3 \pm 2.3) \times 10^{19}$~yr~\cite{bnl} (eval. 2006), which is relatively well reproduced by shell model calculations accounting only for one major shell~\cite{07horoi}.
Meanwhile, according to the experiments by Yako {\it et al.} \cite{09yako}, shell model calculations including only one major shell \cite{07horoi} possibly underestimate the Gamow-Teller transition strength from $^{48}$Ca to $^{48}$Sc and that from $^{48}$Sc to $^{48}$Ti (Fig.~\ref{fig1}).
Note that the transition strength shown in \cite{09yako} may or may not include isovector spin monopole transition strength~\cite{yakonotice} in addition to the Gamow-Teller transition strength.

The ultimate goal of our research project is to present the matrix element of neutrino-less double beta decay process with the highest accuracy so far, and to predict both possible half-life of neutrino-less double beta process and the mass of neutrino.
Among others here we concentrate on establishing a framework of shell model calculation (i.e. the effective nuclear force), which well describes Gamow-Teller transition processes around $^{48}$Ca. 

\section{Nuclear matrix elements of two-neutrino double beta-decay}
We consider two-neutrino double-beta decay process:
\[  ^{48}{\rm Ca}   \to  ^{48}{\rm Ti} + 2 e^- + 2 {\bar \nu},  \]
where $^{48}{\rm Ca}$ and $^{48}{\rm Ti}$ correspond to the initial
and final nuclei respectively.
In particular $^{48}{\rm Sc}$ plays a role of providing intermediate virtual states.
The inverse of the half-life is represented by
\begin{equation}  \label{trate}  [T^{1/2}]^{-1} = G^{2 \nu} |M^{2 \nu}({\rm GT})|^2,   \end{equation}
where $G^{2 \nu}$ is the phase space factor, and $M^{2 \nu}({\rm GT})$ denotes the nuclear matrix elements due to the Gamow-Teller (GT) transition.
Note that transitions other than the GT transition are negligible, as far as two-neutrino double-beta decay processes are concerned~\cite{ejiri}.
Although the value of $G^{2 \nu}$ has not been fixed so far (e.g., see \cite{12kotila}), here we take  $G^{2 \nu} = 1.044 \times 10^{-17}$~yr$^{-1}$~MeV$^2$~\cite{98suhonen}.
This value corresponds to the value adopted in relevant papers~\cite{07horoi,09yako}.
The nuclear matrix element is represented by
\begin{equation} \label{mat1} \begin{array}{ll}
M^{2 \nu}({\rm GT})
 = {\displaystyle \sum_{c=1}^{c_{\max}}} \frac{<0_f^+||(\tau \sigma)^-||1_c^+><1_c^+||(\tau \sigma)^-||0_i^+>}{E_c -E_i +Q_{\beta \beta}/2} 
\end{array} \end{equation}
where states $|0_i>$, $|0_f>$ and $|1_c>$ stand for the $0^+$ ground state of the initial nucleus with the energy $E_i$, $0^+$ ground state of the final nucleus with the energy $E_f$, and $1^+$ intermediate virtual states with the energy $E_c$, respectively.
$Q_{\beta \beta} = E_i - E_f$ denotes the $Q$-value of the double beta decay, and the value of $Q_{\beta \beta}$ is almost precisely determined in recent experiments (e.g., $Q_{\beta \beta} = 4.26698(38)~{\rm MeV}$ \cite{AME2012}). 
The operator $(\tau \sigma)^{\pm}$ means the GT$^{\pm}$ transition
operator, and it is replaced by the effective GT$^{\pm}$ transition operator $(\tau \sigma)^{\pm}_{\rm eff} = q (\tau \sigma)^{\pm} = 0.77 (\tau \sigma)^{\pm}$ in shell model calculations~\cite{90brown}.
Even though a constant $c_{\max}$ is equal to $\infty$ in rigorous treatments, it is replaced by a finite value if the transition through $1_{c ~ (>  c_{\max})}^+$ state is negligible.  
Using the experimental half-life ($(4.3 \pm 2.3) \times 10^{19}$~yr~\cite{bnl}) and Eq.~\eqref{trate}, the matrix element for two-neutrino double-beta decay of $^{48}$Ca is
\[M^{2 \nu}({\rm GT}) = 0.0560 \pm 0.0162~{\rm MeV}^{-1}, \]
while the corresponding value obtained by shell model calculation accounting only for one major shell is 0.0539~MeV$^{-1}$~\cite{07horoi}.
The nuclear matrix element is seemingly well reproduced by a shell model calculation.
However it is notable that cancellations may lead to the unexpected coincidence between experimental and theoretical values for the nuclear matrix element, since the sign of numerator in Eq.~\eqref{mat1} for a certain $c$ is not necessarily positive.
The claim of the discrepancy in GT transition strength~\cite{09yako} implies the reality of such unexpected cancellations.

\begin{figure} [t]
 \includegraphics[width=15.0cm]{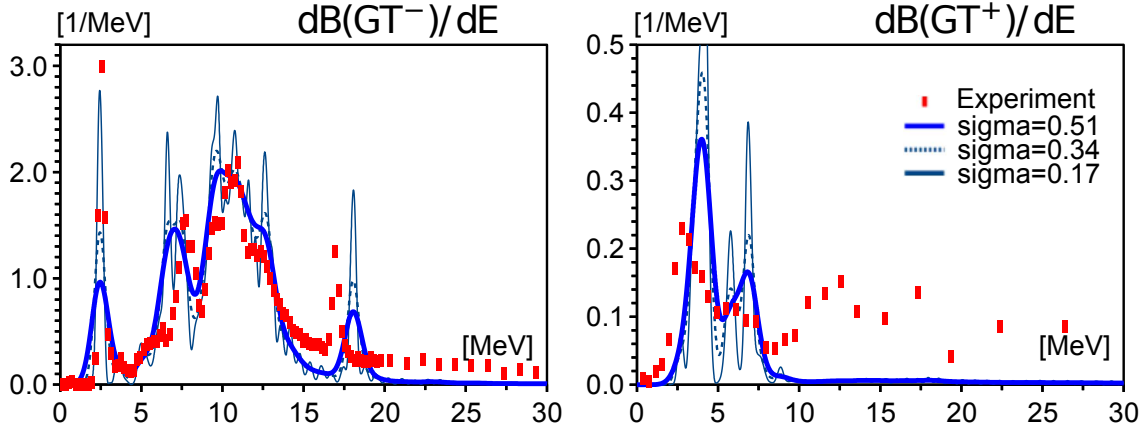}
\caption{(color online) GT$^{\pm}$ transition strength based on shell model calculations (blue curves) accounting for two major shells (SDPFMU-2$\hbar \omega$) and experimental data (red bars)~\cite{09yako}.
Description manner follows from Fig.\ref{fig1}.
} 
\label{fig2}
\end{figure}

\section{Results}
\subsection{Shell model calculations including one major shell}
The GT transition strengths $dB({\rm GT}^{-})/dE$ and $dB({\rm GT}^{+})/dE$ are essentially regarded as the main ingredients of the nuclear matrix element (cf. the numerator in Eq.~\eqref{mat1}):
\[  \begin{array}{ll}
B({\rm GT}^{-};c) = |<1_c^+||(\tau \sigma)^{-} ||0_i^+>|^2, \vspace{2.5mm} \\
B({\rm GT}^{+};c) = |<0_f^+||(\tau \sigma)^{+} ||1_c^+>|^2.
\end{array} \]
A shell model calculation including only one-major shell (employing GXPF1A~\cite{05honma}) is compared with the GT transition experiments in Fig.~\ref{fig1}.
Remarkable discrepancies are noticed in $dB({\rm GT}^-)/dE$ for energies $E^*>12.5$~MeV and in $dB({\rm GT}^+)/dE$ for energies $E^*> 7.5$~MeV respectively, where $E^{*}$ denotes the excitation energy of $1^+_c$ state of $^{48}$Sc measured from the ground state of $^{48}$Sc.
That is, we see no significant difference in low energies satisfying $E^* <  5.0$~MeV, where note that a larger difference in low energies can be found in $dB({\rm GT}^+)/dE$ values compared to $dB({\rm GT}^-)/dE$ values.
Among several reasons for the discrepancy in higher energies, the most crucial
missing contribution is expected to arise from the excitations across the major shells.

\subsection{Shell model calculations including two major shells}
By employing SDPF-MU interaction~\cite{utsuno}, shell model calculations including two major shells ($sdpf$-shell consisting of $sd$ and $pf$ shells) are carried out.
Because of the limited computational power, it is impossible to fully
take into account all the configurations allowed in the $sdpf$-shell model
space (the corresponding $m$-scheme dimension for the diagonalization $>10^{10}$).
Therefore, after employing the Lanczos strength function
method~\cite{caurier}, we truncate the model space in a reasonable manner (the corresponding $m$-scheme dimension $\sim 10^{9}$) in which the excitation is limited up to 2$\hbar \omega$ type excitations (SDPFMU-2$\hbar \omega$).
A shell model calculation including two major shell is compared with the GT transition experiments in Fig.~\ref{fig2}.
The discrepancy between experimental and theoretical values becomes smaller mainly for energy region 12.5$ < E^* <$15.0~MeV of GT$^-$ transition, but there are still significant discrepancies in high energy regions $E^* > 7.5$~MeV of GT$^+$ transition.

\begin{table}[t] \label{table1}
\caption{Proton and neutron excitations across the major shells; the number of excited protons and neutrons (in this order) from $sd$-shell to $pf$-shell are shown in each cell.}
\begin{center}
  \begin{tabular}{|l||c|c||c|c||c|c||} \hline
     & $^{48}$Ca~($0^+_1$) & $^{48}$Ca~($0^+_2$)  & $^{48}$Ti~($0^+_1$) & $^{48}$Ti~($0^+_2$) & $^{48}$Sc~($1^+_{1}$) & $^{48}$Sc~($1^+_{10}$)   \\ \hline \hline
   SDPFMU-2$\hbar \omega$ &  0.17, 0.11 & 0.22, 0.19  & 0.21, 0.19 & 0.20, 0.16 & 0.18, 0.14  &  0.21, 0.17  \\ \hline
  \end{tabular}
\end{center}
\end{table}

\section{Concluding remark}
We have obtained a better description of Gamow-Teller transitions around $^{48}$Ca by introducing the two major shells, but there are still missing higher energy contributions.
The numbers of excited nucleons across the major shells are smaller than expected (Table I); indeed $^{48}$Ca~($0^+_2$) has been claimed to be proton-excitation state~\cite{86videbaek}, so that the corresponding number of excited proton is expected to be close to 2.00.
One reason is that the interaction (SDPF-MU) is not sufficient to describe GT transitions around $^{48}$Ca within the truncated $sdpf$-model space.
In fact the excitation energies of $^{48}$Ca~($0^+_2$) and $^{48}$Ti~($0^+_2$) are 5.099 and 4.097~MeV for SDPFMU-2$\hbar \omega$ calculations, while the corresponding experimental values are 4.283 and 2.997 MeV respectively~\cite{nudat}.  
Note that the corresponding values for GXPF1A calculations are 5.275 and 4.048~MeV respectively.
Further investigation of nuclear interaction is in progress for the better description. \\ \\
This work was supported by HPCI Strategic Programs for Innovative Research Field 5 ``The origin of matter and the universe".
Authors are grateful to Prof. Yako for fruitful comments.


\begin{thebibliography}{9}
\bibitem{yakonotice}
Private communication with Prof. K. Yako; the isovector spin monopole
transition is expected to be included in high energy region.
\end{thebibliography}
\end{document}